\documentclass[pra, aps, showpacs, groupedaddress, superscriptaddress, twocolumn, 10pt]{revtex4-1}

\usepackage{subfigure}
\usepackage[]{graphicx}
\usepackage{textcomp} 
\usepackage{color}
\usepackage{hyperref}
\usepackage{siunitx}
\usepackage{amssymb}
\usepackage{amsmath}

\begin{document}
\title{Extreme Event Statistics in a Drifting Markov Chain}

\author{Farina Kindermann}
\affiliation{Department of Physics and Research Center OPTIMAS, University of Kaiserslautern, Germany}

\author{Michael Hohmann}
\affiliation{Department of Physics and Research Center OPTIMAS, University of Kaiserslautern, Germany}

\author{Tobias Lausch}
\affiliation{Department of Physics and Research Center OPTIMAS, University of Kaiserslautern, Germany}

\author{Daniel Mayer}
\affiliation{Department of Physics and Research Center OPTIMAS, University of Kaiserslautern, Germany}
\affiliation{Graduate School Materials Science in Mainz, Gottlieb-Daimler-Strasse 47, 67663 Kaiserslautern, Germany}

\author{Felix Schmidt}
\affiliation{Department of Physics and Research Center OPTIMAS, University of Kaiserslautern, Germany}
\affiliation{Graduate School Materials Science in Mainz, Gottlieb-Daimler-Strasse 47, 67663 Kaiserslautern, Germany}

\author{Artur Widera}
\affiliation{Department of Physics and Research Center OPTIMAS, University of Kaiserslautern, Germany}
\affiliation{Graduate School Materials Science in Mainz, Gottlieb-Daimler-Strasse 47, 67663 Kaiserslautern, Germany}

\begin{abstract}
We analyse extreme event statistics of experimentally realized Markov chains with various drifts. Our Markov chains are individual trajectories of a single atom diffusing in a one dimensional periodic potential.
Based on more than 500 individual atomic traces we verify the applicability of the Sparre Andersen theorem to our system despite the presence of a drift.
We present detailed analysis of four different rare event statistics for our system: the distributions of extreme values, of record values, of extreme value occurrence in the chain, and of the number of records in the chain. 
We observe that for our data the shape of the extreme event distributions is dominated by the underlying exponential distance distribution extracted from the atomic traces.
Furthermore, we find that even small drifts influence the statistics of extreme events and record values, which is supported by numerical simulations, and we identify cases in which the drift can be determined without information about the underlying random variable distributions. 
Our results facilitate the use of extreme event statistics as a signal for small drifts in correlated trajectories.
\end{abstract}
\maketitle

\section{Introduction}
Extreme events can have profound impact on almost all event series in our lives, ranging from extreme stock market prices, extreme weather events, or the rate of chemical reactions\cite{Sabir2014,Redner2006,Redner2001}.
Due to their immediate relevance such events have been intensely studied on a mathematical level.
A maximum (minimum) in the time series of $n$ entries $\{\xi_0,\xi_1,\ldots, \xi_{n-1}, \xi_n\}$ occurs at step $k$, when $\xi_k$ exceeds (is below) the values of any other entry of the series.
The extreme events, i.e. maxima and minima, are special cases of records. An upper (lower) record is achieved, when $\xi_k$ has a larger (smaller) value than all entries $\xi_l,$ before $ \, l<k$. 
Thus the maximum (minimum) is always the last, and hence highest (lowest), record occurring in the series.

\begin{figure*}
\includegraphics[width=0.95\textwidth]{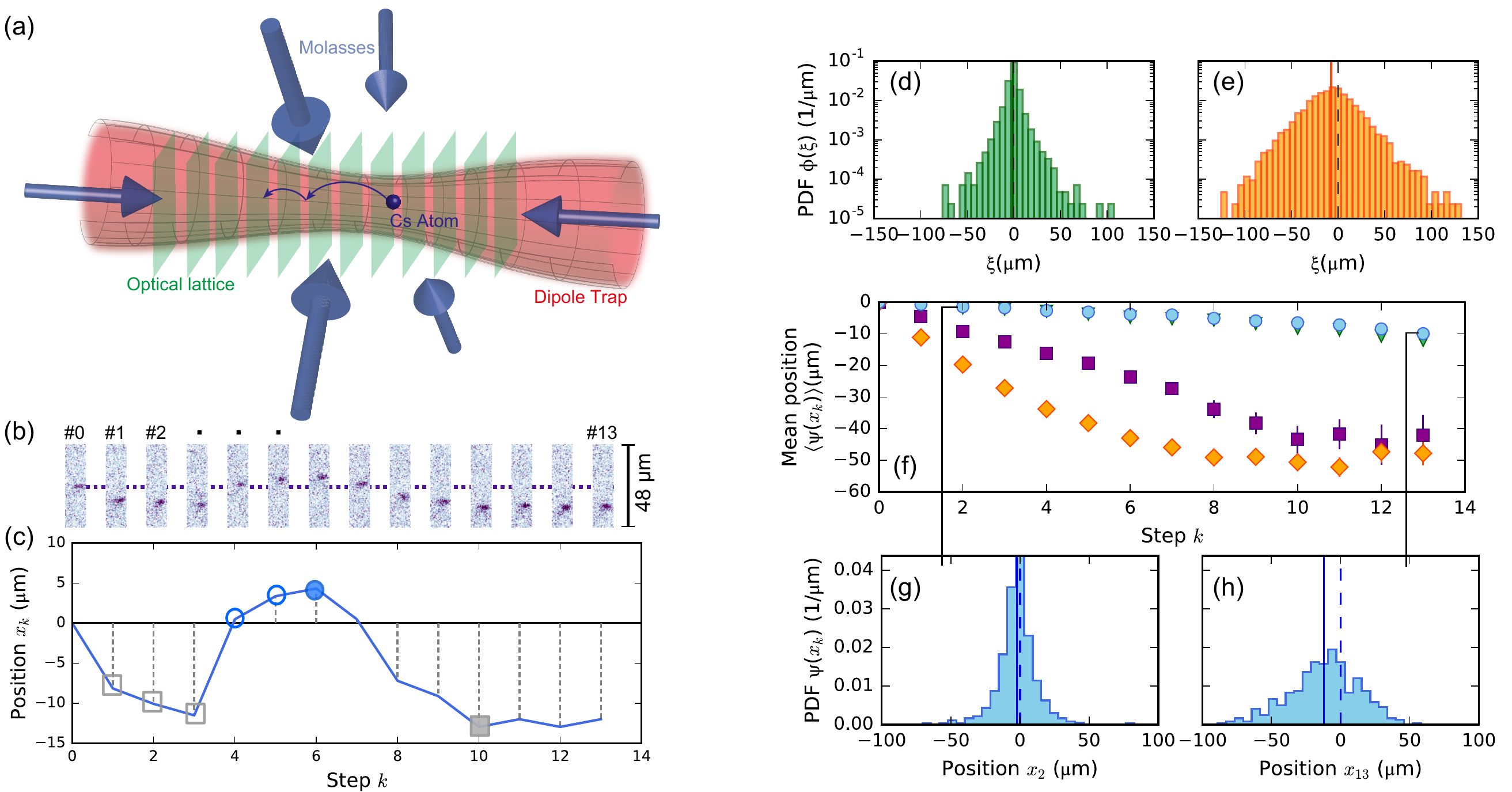}
\caption{Overview of the experiment and its related quantities. (a) Schematic view of the experimental setup. A Cs atom is trapped in a one-dimensional blue-detuned optical lattice, which is overlapped with an optical dipole trap for radial confinement. The diffusion of the atom is initiated by six beams forming a 3D optical molasses. This light field is also used for fluorescence imaging of the atoms to infer their position in the optical lattice. (b) A typical fluorescence trace of the diffusing atom consisting of 14 individual images. The corresponding position trace is depicted in (c). Here, squares (circles) depict lower (upper) records of the trace and the filled symbols indicate the extreme values obtained. Out of hundreds of such traces, we can calculate on one hand the distance distribution $\phi(\xi)$ for different flight times $t_\mathrm{flight}$, shown exemplarily for (d) $t_\mathrm{flight} = \SI{1}{\milli \second}$ and (e) for $t_\mathrm{flight} = \SI{50}{\milli \second}$. While for small $t_\mathrm{flight}$ almost no shift of the mean value (solid vertical lines) from the origin (dashed lines) is detected in the distance distribution, it becomes visible for larger $t_\mathrm{flight}$. (f) Mean values of the position distribution $\langle \psi(x_k) \rangle$ for increasing step $k$ for $t_\mathrm{flight}=\SI{0.1}{\milli \second}$ (blue $\circ$), $t_\mathrm{flight)}= \SI{1}{\milli \second}$ (green $\triangle$), $t_\mathrm{flight)}= \SI{10}{\milli \second}$ (purple $\square$) and $t_\mathrm{flight)}= \SI{50}{\milli \second}$ (orange~$\diamond$). (g) Position distributions $\psi(x_k)$ for $t_\mathrm{flight} = \SI{0.1}{\milli \second}$ with $k = 2$  and (h) $k = 13$, illustrating the increasing importance of the drift with increasing step number $k$. The dashed lines in (g) and (h) again indicate the origin, while solid vertical lines depict the mean values of the position distributions.}
\label{fig:Trace}
\end{figure*}

The statistics of records has evolved over the last decades to a substantial part of probability theory \cite{Barry1998}. Much is known in the case, when the entries $\xi$ of the series are so called independent and identically distributed (i.i.d) random variables (RV) drawn from a distribution $\phi(\xi)$ \cite{Foster1954,Glick1978}.
Such models have been successfully applied for example in sports \cite{Ben-Naim2007,Gembris2002}, biological evolution \cite{Krug2007}, theory of spin glasses \cite{Sibani2006}, or to quantify properties of quantum systems \cite{Bhosale2012,Shashi2013}.
In all of these systems, the entries are completely uncorrelated. However, correlated systems occur often in nature.
A prominent example is the concept of Markov chains, where the entries of the series $x_k$ are correlated through $x_k = x_{k-1} +\xi_k$, yielding a new distribution $\psi(x_k)$.
Here the distances between each step $\xi_k$ are again uncorrelated i.i.d random variables.
A frequently used and powerful model of this concept is the random walk, where positions of the walker constitute the Markov correlated variables. For a random walk in position space, the distribution $\phi(\xi)$ is the distribution of hopping distances $\xi$ in each step, while the distributions $\psi(x_k)$ yields the probability distribution of finding the walker at position $x_k$ in step $k$, see for example Fig.~\ref{fig:Trace}.
Random walk models in general have been successfully applied in physics, biology and economical studies \cite{Denk1989,SchweitzerFrank2007,vandenEngh1992,Bouchaud2003}, and record statistics for continuous-time random walks have been investigated \cite{Sabhapandit2011}.

Based on individual trajectories of a walker, one can distinguish four different extreme event distributions:
(i) the probability distribution of the maximal and minimal values of each trajectory and likewise (ii) the record value distribution; (iii) the probability distribution at which step $k$ in the sequence a maximum or minimum occurs, and finally (iv) the distribution of record numbers within each trajectory. Finding analytical solutions for such statistics is  challenging for Markov correlated events and thus interest in this field has grown recently \cite{Majumdar2010b}.

The walk underlying the Markov process is often complicated by an additional drift $\mu$ added in each step, thus the position in step $k$ reads $x_k = x_{k-1}+\xi_k+\mu_k$, where in the simplest case a constant drift $\mu_k=\mu$ is applied.
Adding the drift leads to a position distribution $\psi(x_k)$ constantly shifted away from the origin. Thus all extreme event distributions mentioned above are expected to be also affected by such a drift.
While analytical predictions on record statistics have been made \cite{LeDoussal2009,Majumdar2012}, only little is known about the extreme events in this case.
It has been shown that, e.g., cash flows of insurance companies \cite{Gerber1996,Dickson2001}, the energy dissipation of optical beams in self-defocusing media \cite{Villarroel2010} or the dispersive transport of contaminant particles \cite{Edery2011} are successfully modelled by Markov chains undergoing a drift.

One of the central questions of such systems is how much information on the drift can be extracted from knowledge of extreme events only, if, e.g., only record numbers or maximal values are known while the underlying distributions are (partially) unknown. An additional motivation can arise, where large data sets need to be analysed, and where an important question is if equivalent results can be obtained from a reduced set of data. 

We approach these questions by comparing the information obtained from the fundamental distributions $\phi(\xi)$ and $\psi(x_k)$ on the one hand with the four distributions of extreme events and records on the other hand. 
\par
Experimentally, we generate Markov chains including a drift using a single Caesium (Cs) atom diffusing in a periodic potential due to driving by an external light field. 
By stroboscopically interrupting the diffusion process and imaging the atomic position with fluorescence imaging, we have access to the individual trajectories, see Fig. \ref{fig:Trace}. The extreme event statistics of this system is interesting itself, as it has been shown to feature an exponential distance distribution $\psi(\xi)$ combined with a slow relaxation to ergodic behavior: For all time scales observed the single trajectory is not representative for the ensemble of particles, see Ref.~\cite{Kindermann2016}. Employing such a tightly controllable experimental system has the additional advantage that it ensures statements that are robust against noise. In fact, noise and measurement errors have been shown to have a strong influence onto record statistics \cite{Edery2013}.  We support our findings with numerical simulations based on i.i.d RV for the distribution $\phi(\xi)$. 

We find deviations from the expected behaviour for systems without drift, thus a clear signature of the drift can be derived from extreme event statistics. In particular, we find that, for our system, the shape of the distance distribution $\psi(\xi)$ prevails for extreme event distributions, in contrast to systems without drift. Furthermore we conclude that full information on the drift without knowledge about the underlying distributions can only be obtained from considering extreme event values.  

In section \ref{sec:Experiment} the experimental setup and data obtained are discussed. We then test the applicability of the Sparre Andersen theorem to our data in section \ref{sec:SparreAndersen}. In section \ref{sec:ValueDistributions} we analyze the distributions of extreme values and records, and consider the occurrance of extreme events in section \ref{sec:Occurance} before we analyze the record number statistics in section \ref{sec:RecordNumbers}.

\section{Experimental setup} \label{sec:Experiment}
In order to investigate the statistics of extreme events and records, we use a single Cs atom diffusing in a periodic potential, driven by a near resonant light field.
To this end, we capture atoms in a high magnetic field gradient magneto optical trap (MOT). The atom number observed shows a Poissonian probability distribution with unity as the most probable value, and we postselect realisations with one atom for the following analysis.
Subsequently, the atom is loaded into a periodic potential, i.e. an optical lattice, at lattice depth $U_0 = k_\mathrm{B} \times \SI{850}{\micro \K}$, with $k_\mathrm{B}$ the Boltzmann constant. 
The lattice is formed by two counter propagating laser beams at a wavelength of $\lambda_\mathrm{lat} = \SI{790}{\nano \m}$ and blue detuned to the Cs D2 line. Hence, atoms are only confined along the lattice axis.
To also provide radial confinement, the lattice beams are spatially superposed with a running wave laser beam at $\lambda_\mathrm{DT} = \SI{1064}{\nano \m}$ forming a dipole potential with depth $U_\mathrm{DT} = k_\mathrm{B} \times \SI{1}{\milli K}$. All detunings of the optical potential from the Cs transitions are large enough to neglect photon scattering on relevant time scales of the experiment.
In the lattice we take a fluorescence image with exposure time of \SI{500}{\milli \second}, from which we deduce the initial atomic position.
Subsequently, the periodic potential is lowered in \SI{10}{\milli \s} to a value $U_\mathrm{low} = k_\mathrm{B} \times \SI{210}{\micro \K}$, while the dipole trap potential is held constant, to allow diffusion along the lattice axis only.
The diffusion process is initiated by illuminating the atom by an optical molasses. It is formed by three mutually orthogonal pairs of counter propagating laser beams at a total power of \SI{500}{\micro \W} and a red detuning of $\Delta\approx 1 \times \Gamma_\mathrm{nat}$, with $\Gamma_\mathrm{nat} = 2\pi \times \SI{5.22}{\mega \Hz}$ \cite{Steck2008} the natural linewidth of the Cs $D_2$ transition.
The optical molasses on the one hand acts as a random force by constantly scattering photons at rate $\Gamma_\mathrm{phot} = \SI{2e6}{\per \s}$.
On the other hand, the atoms are constantly cooled by laser cooling, resulting in a temperature of $T_\mathrm{Cs} \approx \SI{50}{\micro \K}$.
After a variable diffusion time $t_\mathrm{flight}$ between \SI{0.1}{\milli \s} and \SI{50}{\milli \s}, the lattice potential suddenly is increased to $U_0$ again, such that the diffusion process is interrupted and the atomic position is effectively frozen. 
We take another fluorescence image, from which we deduce the new atomic position $x_k$ and thus the total distance travelled $\xi_k$ by the atom during $t_\mathrm{flight}$.
For each trajectory we repeat this diffusion sequence in total 13 times, resulting in 13 diffusion steps and thus a maximal trace length of $n = 14$ as also the initial position $x_0$ is counted.
A typical trajectory and the respective fluorescence images are depicted in Fig. \ref{fig:Trace}(b,c).
The lattice confinement together with the photon scattering leads to an escape time  from a lattice well of $\tau_\mathrm{esc} =\SI{7}{\milli \second}$ \cite{Hanggi1990,Kindermann2016} and thus allows to observe the atomic trajectories.
\par
Out of these fluorescence images we are further able to extract an exponential hopping distance distribution $\phi(\xi)$. 
For small flight times $t_\mathrm{flight}$ less than one jump occurs on average, while for the longest flight time $t_\mathrm{flight} = \SI{50}{\milli \s}$ the atom might jump more than once. Consequently, the exponential distance distribution $\phi(\xi)$ for individual jumps is reflected by the position distribution $\psi(x_k)$ for small flight times. By contrast, for longer flight times the central limit theorem transforms the position distribution $\psi(x_k)$ for large $k$ into a Gaussian. 
Neglecting the rapid in-well dynamics, our system is well described by a continuous time random walk model with exponential distance and waiting time distributions (for details see  \cite{Kindermann2016}).

Additionally, atoms experience a small drift to \emph{negative} positions, originating from slight misalignments, for example of the power of two counter-propagating molasses beams.
The drift, however, is small compared to the width of the hopping distance distribution $\phi(\xi)$, and for small flight times it is smaller than the resolution of the imaging system which is \SI{2}{\micro \meter}. Thus in a first approximation, this drift is expected to be negligible to describe the dynamics of the walker, see Fig.~\ref{fig:Trace}(d,e).
\begin{table}
\begin{ruledtabular}
\begin{tabular}{c|c|c|c|c}
$t_\mathrm{flight}$ (ms) & 0.1 & 1 & 10 & 50 \\ \hline
$\mu$ (\si{\micro \meter}) & -0.7 & -0.8 & -2.3 & -7.5 \\ \hline
$\sigma$ (\si{\micro \meter}) & 7 & 8 & 15 & 24 \\ \hline
$\mu_\mathrm{rel}$  & 0.1 & 0.1 & 0.15 & 0.31 \\
\end{tabular}
\end{ruledtabular}
\caption{Summary of the flight times used and the corresponding drifts $\mu$ inferred from the mean value of the position distribution and widths of the distance distribution $\sigma$ obtained. The ratio of $\mu$ and $\sigma$ is given by the relative drift $\mu_\mathrm{rel}$. For details see text.} 
\label{tab:Meas}
\end{table}
Experimentally, we are able to choose the drift value via the length of the flight time, because the influence of the mechanism causing the drift increases with increasing $t_\mathrm{flight}$, see Table \ref{tab:Meas}.
Here, we calculate the drift $\mu$ as the mean of the distance distribution comprised of at least 6000 individual distances.
Additionally, the drift is inferred from the shift of the position distribution in each step, see Fig.~\ref{fig:Trace}(f,g,h). When averaging over all step numbers in this case, we find the same statistical uncertainty as is gained from the distance distribution.
Measuring $\mu$ in each step individually, see Fig.~\ref{fig:Trace}(f), shows an approximate linear drift for the first few steps, becoming non-linear for step numbers $k>7$ \footnote{The limited width of the detection region also contributes to the complex drift observed.}; this measurement allows direct comparison of the results gained from the extreme event and record analysis presented in the following.
Due to the increase of the flight distance with increasing $t_\mathrm{flight}$ the distance distribution clearly broadens, too, and thus its width, quantified by the standard deviation $\sigma$ of the distribution, increases.
We therefore also define the relative drift $\mu_\mathrm{rel} = \mu/\sigma$, indicating the influence of the drift compared to the width of the distance distribution.
Therefore each of the data sets measured correspond to a different drift value $\mu$ and width $\sigma$.
The data summarized in Tab.~\ref{tab:Meas} further illustrates that for $t_\mathrm{flight} < \SI{10}{\milli \s}$ the value of $\mu$ is smaller than the width of the time average position of the atom, which is determined from the optical resolution to detect the position $x_k$ in step $k$. 
Thus, the drift becomes prominent in the distance distribution for large $t_\mathrm{flight}$, see also Fig.~\ref{fig:Trace}(d,e). 
\par
We numerically model traces of the atoms by drawing a random value in each step from an initial distribution of iid variables, which have the same form (i.e. first and second moment) as the measured distributions $\phi(\xi)$. By adding up each value step by step and introducing a constant shift, we generate both, individual trajectories and the corresponding position distributions $\psi(x_k)$. Based on that data we can then evaluate the extreme value and record statistics in the same way as on the data measured.

\section{Sparre Andersen Theorem}
\label{sec:SparreAndersen}
Analysis of extreme events and particularly universal conclusions drawn rely on a fundamental and universal concept of first passage processes, quantified by the Sparre-Andersen theorem.
\begin{figure}
\includegraphics[width=0.43\textwidth]{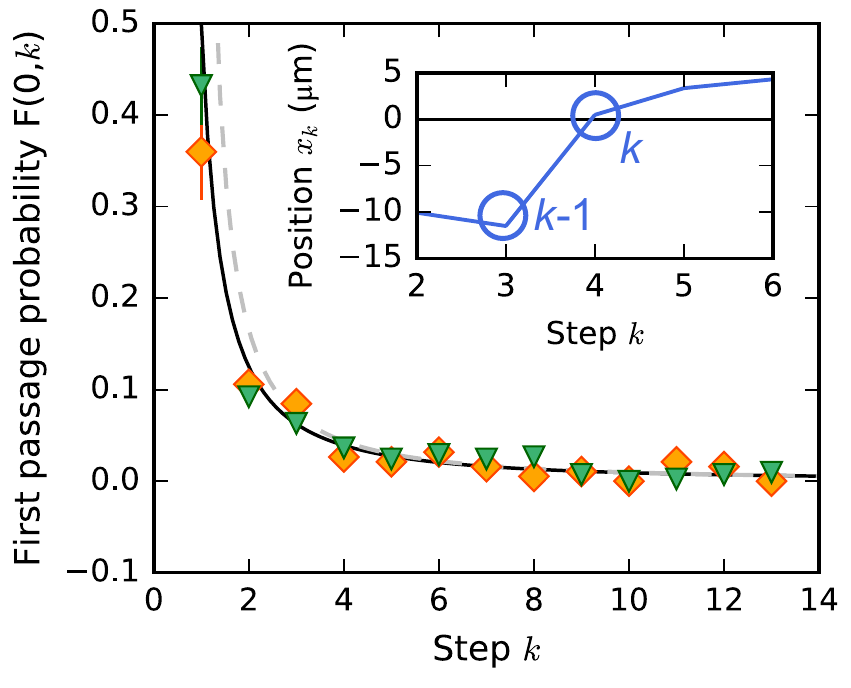}
\caption{First passage time probability at step $k$ for an atom starting its random walk at the origin $x_0 = 0$. The inset depicts the first passage from the negative to the positive semi-axis of the atom of which the full trajectory is shown in Fig.~\ref{fig:Trace}c. Data points are for $t_\mathrm{flight} = \SI{1}{\milli \s}$ (green triangles) and $t_\mathrm{flight} = \SI{50}{\milli \s}$ (orange diamonds). The probability is given by the Sparre Andersen formula (see Eq. \ref{eq:FPT}) depicted as the black solid line, which fits equally well to both data sets. The dashed gray line depicts the large $k$ limit. Error bars are standard statistical errors.
}
\label{fig:sparrAnd}
\end{figure}
In the 1950's Sparre Andersen calculated the probability for a random walker starting at the origin to stay at the positive semi-axis until step~$k$, i.e. the survival probability  
\begin{equation}
Q(x_0 = 0,k) = \binom{2k}{k} 2^{-2k}
\end{equation}
to be independent of the initial distribution $\phi(\xi)$ of displacements, thus being universal, as long as $\phi(\xi)$ is continuous and symmetric \cite{SparreAndersen1954,Majumdar2010b}.
The theorem is valid for all step numbers $k$, but there exists a simple, large $k$ approximation $Q(x_0 = 0,k) \propto 1/\sqrt{\pi k}$ for the survival probability, which for our data seems to apply already for steps larger than three, cf. Fig.~\ref{fig:sparrAnd}.  
It is worth noticing that the universality only holds for particles starting at the origin $x_0 = 0$ and is lost for an arbitrary starting point $x_0 \neq 0$.
A generalization to starting points $x_0 > 0$ was found by Pollaczek and Spitzer \cite{Pollaczek1952,Spitzer1956}. 
However, we define the starting point of each atom as the origin without loss of generality.
From the Sparre Andersen theorem one can calculate the first passage probability
\begin{equation}
\label{eq:FPT}
F(0,k) = Q(x_0 = 0, k-1)-Q(x_0 = 0,k),
\end{equation} 
i.e.~the probability that the walker is on the positive semi-axis up to step $k-1$ but located on the negative semi-axis in step $k$, see inset Fig. \ref{fig:sparrAnd}. 
Although our hopping distance distribution is quasi-discrete (on the scale of the lattice spacing) and a drift is present, the Sparre Andersen formula well applies to the measured probability and that is furthermore true for all drift values measured, see Fig. \ref{fig:sparrAnd} showing the two extreme cases.
For the initial step one would expect a probability of 0.5, as the atom (or walker) either jumps to the positive or negative semi-axis and thus is either counted as a first passage event in step $k=1$ or not.
Due to the drift present in the system, however, it is more likely for the atom to hop to the negative than to the positive side in the initial step.
Therefore the first passage probability (as shown in Fig. \ref{fig:sparrAnd}) is lowered for the initial step.
The good agreement of the rest of the data points with theory
motivates modelling our data using theory based on the Sparre Andersen theorem.
\begin{figure}[b]
\includegraphics[width=0.42\textwidth]{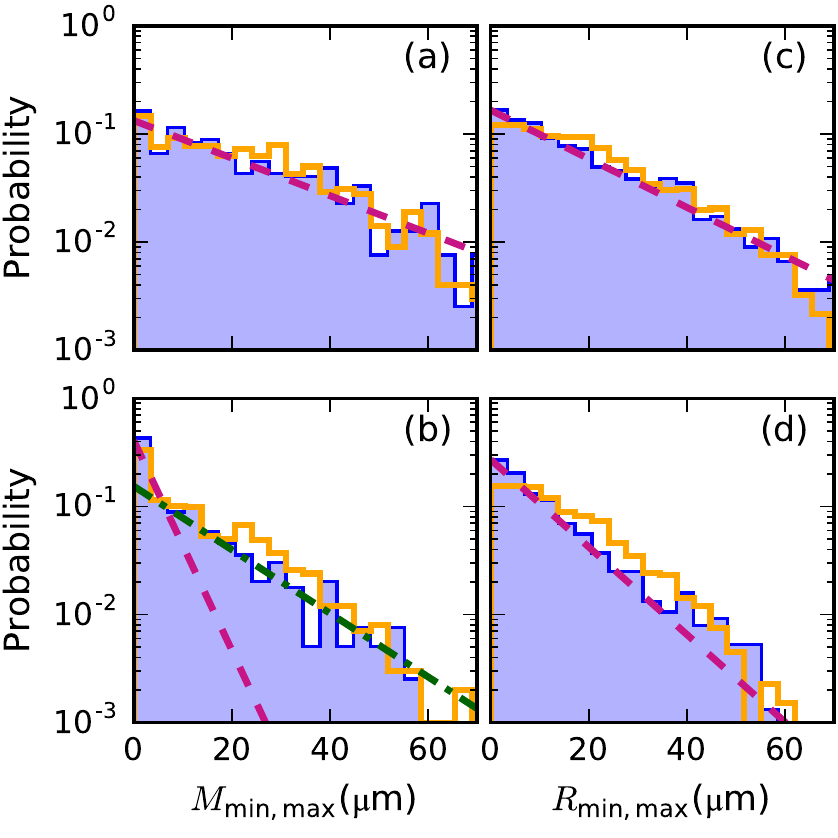}
\caption{Measured (blue, filled) distributions of $M_\mathrm{min}$ (a), $M_\mathrm{max}$ (b), $R_\mathrm{min}$ (c) and $R_\mathrm{max}$ (d). The solid (orange) line indicates the result of numerical simulations in good agreement with the measured data and the dashed (red) line is an exponential fit to the data. Because the atom is more likely to cross the origin directly in the beginning for a system with drift, as reflected by the first passage probability (see fig. \ref{fig:sparrAnd}), $M_\mathrm{max}$ experiences a large peak at zero. Therefore the fit does not perfectly resemble the distribution. However, when excluding the zero peak, the data is still described by an exponential distribution, see dash-dotted(green) line.
}
\label{fig:DistEVS}
\end{figure}
\section{Extreme Value and Record Value Distributions}
\label{sec:ValueDistributions}
\begin{figure}[b]
\includegraphics[width=0.43\textwidth]{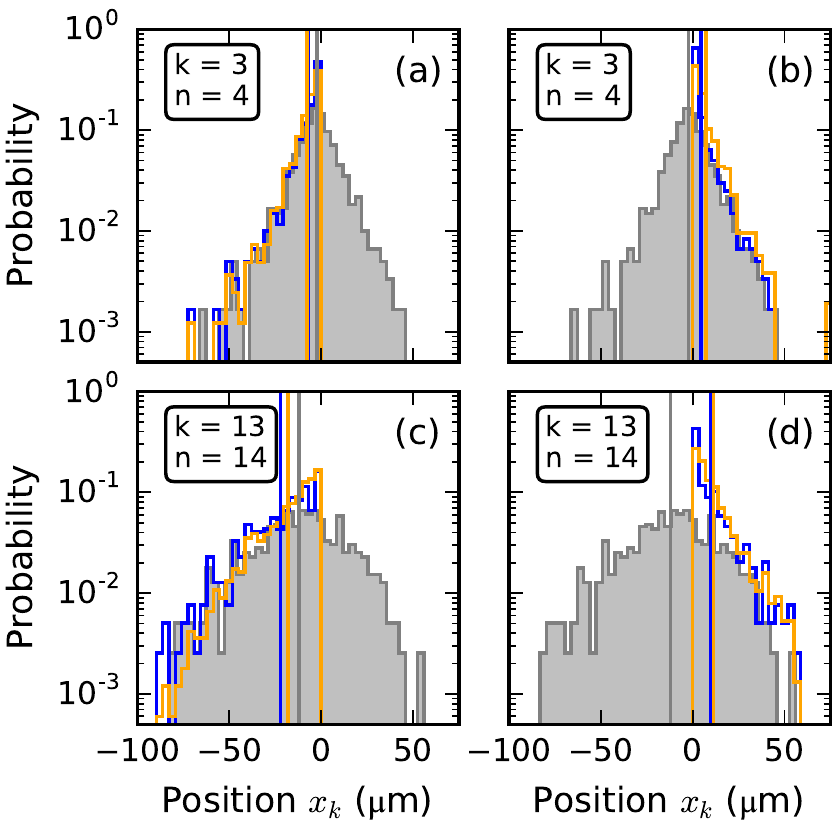}
\caption{Position distribution $\psi(x_k)$ (grey, filled) plotted together with the distribution of $M_\mathrm{min}$ (dark blue solid line) and $R_\mathrm{min}$ (yellow solid line) in (a) and $M_\mathrm{max}$ (dark blue solid line) and $R_\mathrm{max}$ (yellow solid line) in (b), respectively. The same is shown in (c,d), but for a trace length of $n=14$. While the position distribution $\phi(x_k)$ contains all positions an atom occupied in step $k$, the extreme value distribution includes only the maximum (in positive or negative direction) reached by the atom \emph{until} step $k$. The record value distribution is further connected with the extreme value distribution as the last record (value) achieved is always the extreme value. 
Moreover, the first value reached on the negative (positive) side counts as the first lower (upper) record value, as it is always smaller (greater) than zero.
}
\label{fig:DistEVSPos}
\end{figure}
In close relation to the Sparre Andersen theorem is the statistics of extreme values and records, as many properties of these can be derived from that theorem. An example is the distribution of times at which the walker is farthest away from the origin or the number of records achieved in the trace \cite{Majumdar2010b}, as discussed below.
However, calculating the distribution of extreme values or record values from the distance or position distributions remains challenging even for correlated systems without drift \cite{Majumdar2003}.

We first investigate the distributions of maxima ($M_\mathrm{max}$) and minima ($M_\mathrm{max}$) as well as the distributions of upper ($R_\mathrm{min}$) and lower record values ($R_\mathrm{min}$). For systems without drifts and correlations, one would expect one of three distributions~--~Weibull-, Frechet- or Gumbel- distribution -- to describe the case of extreme value statistics. However, we find for the longest step number of $n=14$ an exponential distribution for all distributions and drift values measured, see Fig.~\ref{fig:DistEVS} for an example, suggesting that for our case the distance distribution dominates the extreme event statistics.

An intuitive picture for the effect of drifts on extreme events can be obtained by comparing the nested distributions of positions $\psi(x_k)$ and of extreme and record values, see Fig.~\ref{fig:DistEVSPos}.
Clearly, all records are found in the wings of the position distributions $\psi(x_k)$, while the extreme values comprise only the last record values.
Consequently, extreme value and record value distributions differ from each other with increasing trace length as illustrated in Fig.~\ref{fig:DistEVSPos}. Here, the averages of extreme value distributions show faster dynamics than the position distribution, while the asymmetry indicates the effect of the drift:
For a system without drift $\langle M_\mathrm{min} \rangle$ and $\langle M_\mathrm{max} \rangle$ (and $\langle R_\mathrm{min} \rangle$ and $\langle R_\mathrm{max} \rangle$, respectively) should grow symmetrically.
To obtain a model-independent measure for the drift, we thus calculate the asymmetry of the mean values $\delta_M = \langle M_\mathrm{min} \rangle + \langle M_\mathrm{max} \rangle$ ($  \delta_R =  \langle R_\mathrm{min} \rangle + \langle R_\mathrm{max} \rangle$, respectively) and compare them to  $\langle \psi(x_k) \rangle$. 

\begin{figure}
\includegraphics[width=0.42\textwidth]{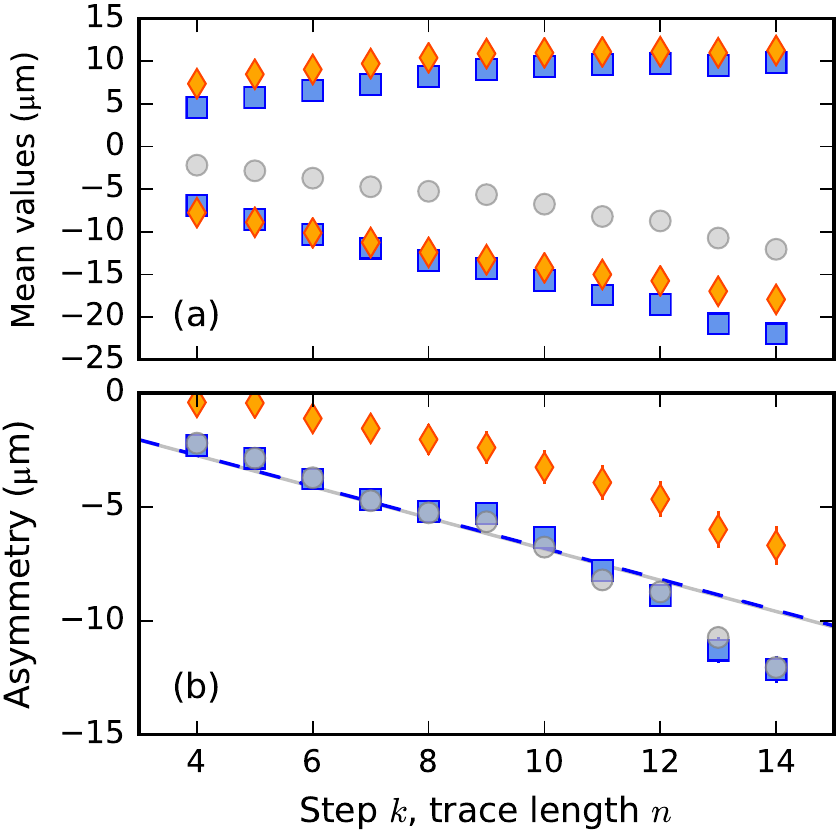}
\caption{(a) Comparison of $\langle M_\mathrm{min,max} \rangle$ (blue $\square$), $\langle R_\mathrm{min,max} \rangle$ (orange $\diamond$) and $\langle \psi(x_k) \rangle$ (light gray $\circ$) for $t_\mathrm{flight} = \SI{1}{\milli \s}$. (b) Differences of the mean of the extreme values distributions $\delta_\mathrm{M}$ and of the mean of the record value distributions $\delta_\mathrm{R}$ for increasing trace length. 
The solid (blue) line is a linear fit to $\delta_\mathrm{M}$ overlapping with a linear fit to $\langle \psi(x_k) \rangle$ (dashed gray line).
}
\label{fig:EVSMeanVals}
\end{figure}
\begin{figure}
\includegraphics[width=0.47\textwidth]{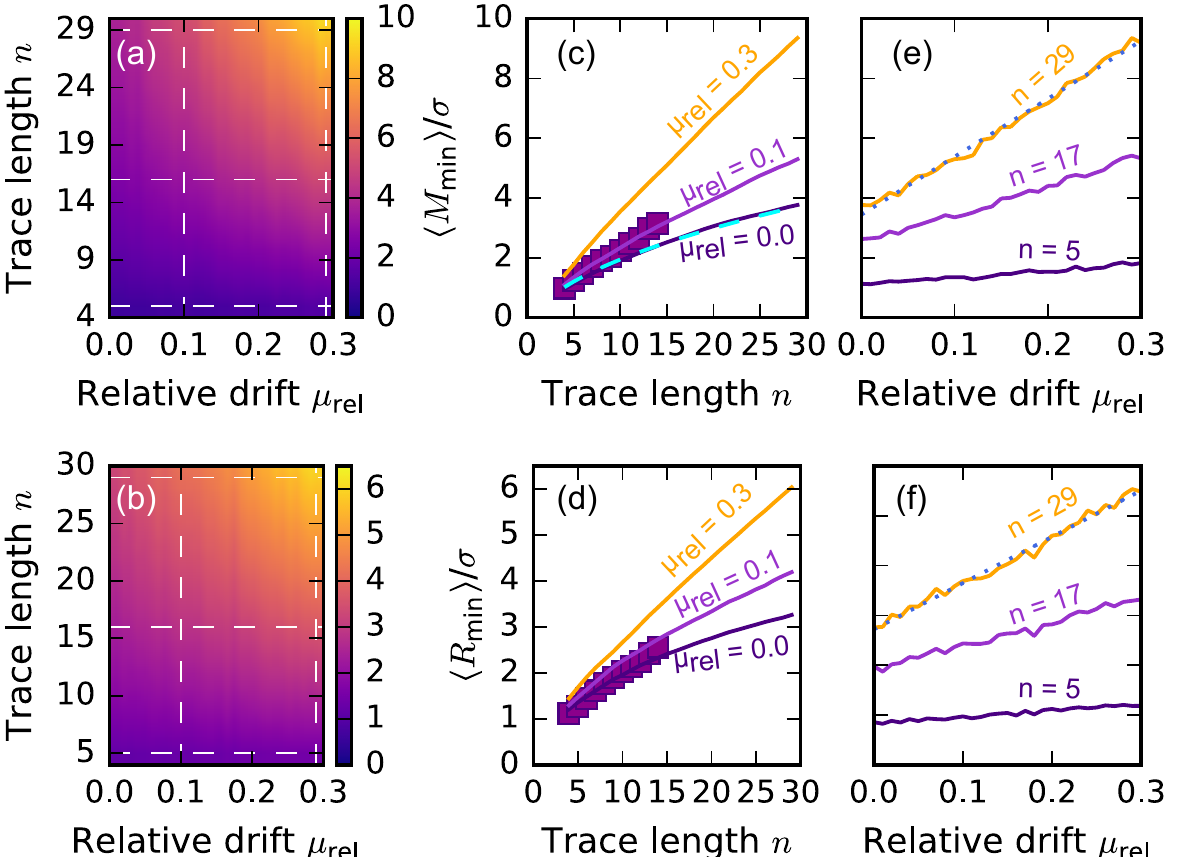}
\caption{(a)Simulated expectation value of $\langle M_\mathrm{min} \rangle/\sigma$ ( color bar)  and of (b) $\langle R_\mathrm{min} \rangle/\sigma$ (color bar) for increasing $\mu_\mathrm{rel}$ and trace length $n$. Dashed vertical lines are cuts at specific $\mu_\mathrm{rel}$ depicted in (c,d) and horizontal dashed lines are cuts at specific trace lengths $n$ depicted in (e,f). The data points are measured values for $\mu_\mathrm{rel} = 0.1$ corresponding to $t_\mathrm{flight} = \SI{1}{\milli \s}$ and the dashed blue line in (c) indicates the theoretical prediction for a system without drift cf. Eq.~\eqref{eq:ExpectMax}. A linear increase is found for $\langle M_\mathrm{min} \rangle/\sigma$ (and $\langle R_\mathrm{min} \rangle/\sigma$, respectively) with increasing $\mu_\mathrm{rel}$, which is indicated by the dotted lines in (e,f).}
\label{fig:EVSExpectVals}
\end{figure}
The resulting $\delta_M$ and $\delta_R$ are depicted in Fig. \ref{fig:EVSMeanVals}b. 
The perfect overlap of $\langle x_k \rangle$ and $\delta_M$ implies that the drift present can be directly calculated from the extreme value statistics without further knowledge of the underlying distribution $\phi(\xi)$, or even $\psi(x_k)$.
Moreover, it demonstrates that the drift $\mu_\mathrm{EV}$ obtained from $\delta_M$ tightly follows the same behaviour as the drift $\mu$ of the position distribution, with  $\mu/\mu_\mathrm{EV} = 1.00 \pm 0.05$ (when averaging over all measurements with various $t_\mathrm{flight}$). 
We conclude that, without any knowledge of the distance or position distribution or related quantities, one is able to infer the drift in the system from the distribution of extreme values only. 

Although the distance distribution may not be known in many cases it is interesting to compare the evolution of $\langle M_\mathrm{min}\rangle$ (and $\langle R_\mathrm{min}\rangle$) with existing predictions.
Already for systems without drift $\langle M_\mathrm{min}\rangle$ is challenging to calculate analytically and strongly depends on the form and width of $\phi(\xi)$ \cite{Majumdar2010b}. It is given by
\begin{align}
\label{eq:ExpectMax}
\langle M_\mathrm{min}(n) \rangle = \sigma \sqrt{\frac{2n}{\pi}}-c,
\end{align}
with $\sigma$ the width of $\psi(\xi)$ and $c$ a constant depending on the form of the distance distribution \cite{Comtet2005,Majumdar2010b}.
When adding a drift we see deviations from this Eq.~\ref{eq:ExpectMax} and find a more linear increase of $\langle M_\mathrm{min}\rangle$ ($\langle R_\mathrm{min}\rangle$) with increasing trace length $n$ in the measurements as well as in numerical simulations, see Fig.~\ref{fig:EVSExpectVals}.
Additionally, the simulations show an approximately linear increase of $\langle M_\mathrm{min}\rangle$ ($\langle R_\mathrm{min}\rangle$) with increasing $\mu_\mathrm{rel}$, which is more and more pronounced for longer trace lengths (see again Fig.~\ref{fig:EVSExpectVals}).
While a comparison of the experimental values with the numerical predictions can also reveal information on the drift, this requires knowledge about the width of the distance distribution $\phi(\xi)$. Thus the model-independent quantity $\delta_M$ is a more directly available measure to infer a drift present in the system.
\section{Occurence of Maxima and Minima}
\label{sec:Occurance}
\begin{figure}
\includegraphics[width=0.41\textwidth]{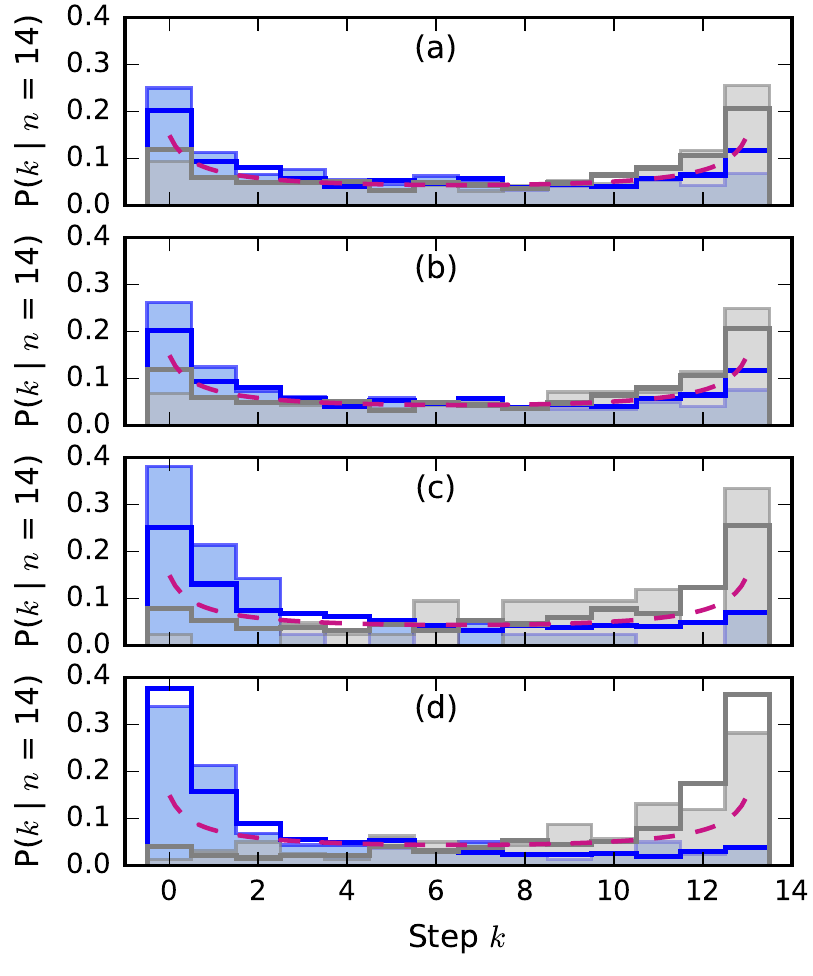}
\caption{Measured occurrences of maxima (dark blue) and minima (light gray) for $t_\mathrm{flight} = \SI{0.1}{\milli \second}, \SI{1}{\milli \second}, \SI{10}{\milli \second},$ and $\SI{50}{\milli \second}$ (a-d) for fixed $n = 14$. Solid lines indicate the results from numerical simulations which are in good agreement with the data measured. The dashed red line indicates the result derived from the Sparre Andersen theorem in Eq.~\eqref{eq:maxOcc}.}
\label{fig:MaxOccurDrift}
\end{figure}
Additionally to the value of an extreme event, another important quantity is the step in the sequence when a maximum or minimum occurs.
This quantity is a random variable and is closely related to the survival probability $Q(0, n)$, i.e. to the Sparre Andersen theorem (for a derivation see \cite{Majumdar2010b}). 
As a consequence it is universal and hence valid for all continuous and symmetric jump distributions.
For a set of trajectories with in total $n$ steps, the probability for the maximum to occur at step $k$ reads~\cite{Majumdar2010b}
\begin{equation}
\label{eq:maxOcc}
P(k|n) = \binom{2k}{k}\binom{2(n-k)}{n-k} 2^{-2n}.
\end{equation}
This is valid for all $n$, rather than being the limiting case for large $n$, and $k$ is again a random variable.
From the experimental data, we investigate the step at which the maximum (minimum) of the trajectory occurs. 
When comparing the results to the theoretical expectation of Eq.~\ref{eq:maxOcc}, a clear deviation is observed, see Fig.~\ref{fig:MaxOccurDrift}.
For a system without drift the maximum (minimum) occurs at the initial or final step with equal probability in contrast to the data measured, where the maximum (minimum) is more likely to prevail at the beginning (end) of the trajectory. 
This is clearly a consequence of the drift, as it is more probable for the atoms to leave the origin in negative direction.
This effect increases for increasing drift (and trace length), but already for a relative drift $\mu_\mathrm{rel} = 0.1$ (at $n = 14$), a large difference is visible, see Fig.~\ref{fig:MaxOccurDrift}(a,b).
\begin{figure}
\includegraphics[width=0.48\textwidth]{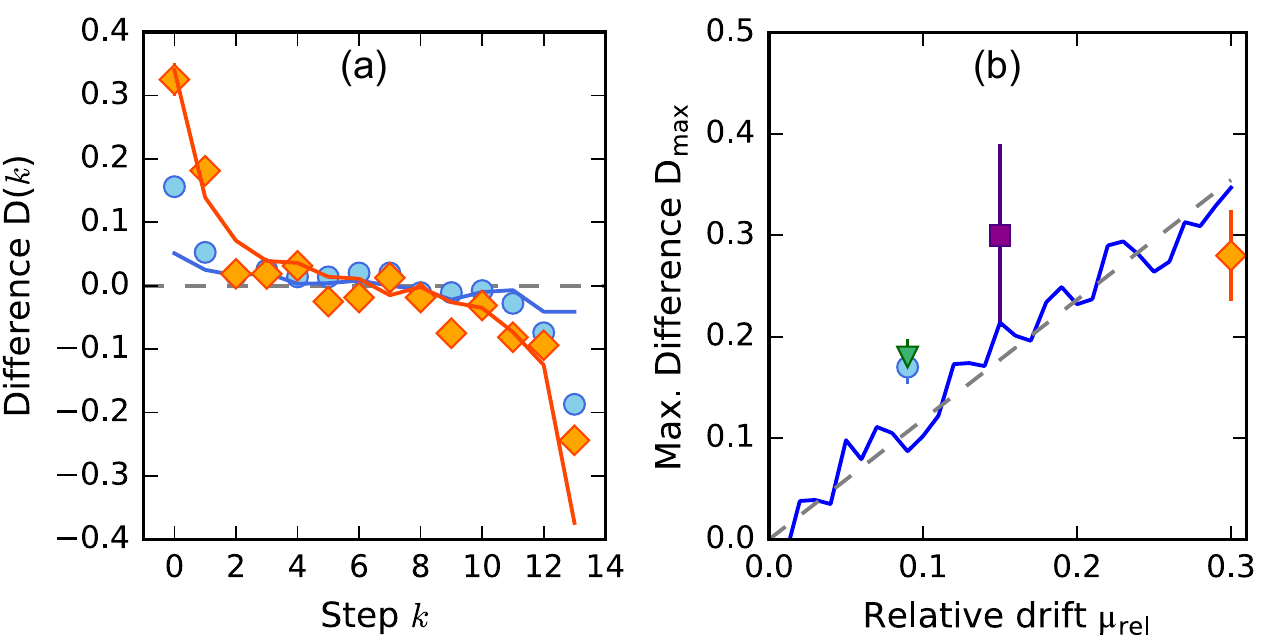}
\caption{(a) Contrast extracted from occurrences of maxima and minima for $t_\mathrm{flight} = \SI{0.1}{\milli \s}$ (blue $\circ$),  and \SI{50}{\milli \s} (orange $\diamond$) and $n = 14$ (as shown in Fig.~\ref{fig:MaxOccurDrift}(b and d)). Solid lines indicate the results from numerical simulations. (b) Data points depict the maximal contrast extracted from the data in (a), while symbols and colors match the ones in Fig.~\ref{fig:Trace}(f). The solid line again indicates the result from numerical simulations and the dashed line depicts a linear fit to the simulation.} 
\label{fig:Contrast}
\end{figure}
To quantify the difference in occurrence of maxima and minima, we calculate the difference $D(k)$, defined as $D(k) = P_\mathrm{max}(k|n) - P_\mathrm{min}(k|n)$, with $P(n|k)$ the measured probability of the maximum (minimum) occurring at step $k$ for a given trace length $n$.
The results for a trace length of $n$ = 14 and $\mu_\mathrm{rel} = 0.1$ and $\mu_\mathrm{rel} = 0.31$ together with the respective results from numerical simulations are depicted in Fig.~\ref{fig:Contrast}(a).
As a measure for the drift, we use the maximal difference $D_\mathrm{max} = (D(0)-|D(13)|)/2$, which we calculate for all drift values at $n = 14$, as depicted in Fig.~\ref{fig:MaxOccurDrift}.
With increasing $\mu_\mathrm{rel}$ $D_\mathrm{max}$ increases, too, and from numerical simulations an approximately linear behaviour is extracted as shown in Fig.~\ref{fig:Contrast}(b).
We attribute the slight discrepancy between the measured data and the simulations to the fact that the simulations overestimate the occurrences of maxima. The simulation assumes a linear drift, while the experimental system shows a more complex drift, see Fig.~\ref{fig:Trace}.
However the simulations and the data both show the same correlation of relative drift with contrast of the occurrence probabilities.
Thus the probability of the step at which the maximum (minimum) appears is a measure for drifts. Importantly, it is applicable for systems, where the underlying distance distribution is not known. 

\section{Record Number Distributions}
\label{sec:RecordNumbers}
%
\begin{figure}
\includegraphics[width=0.42\textwidth]{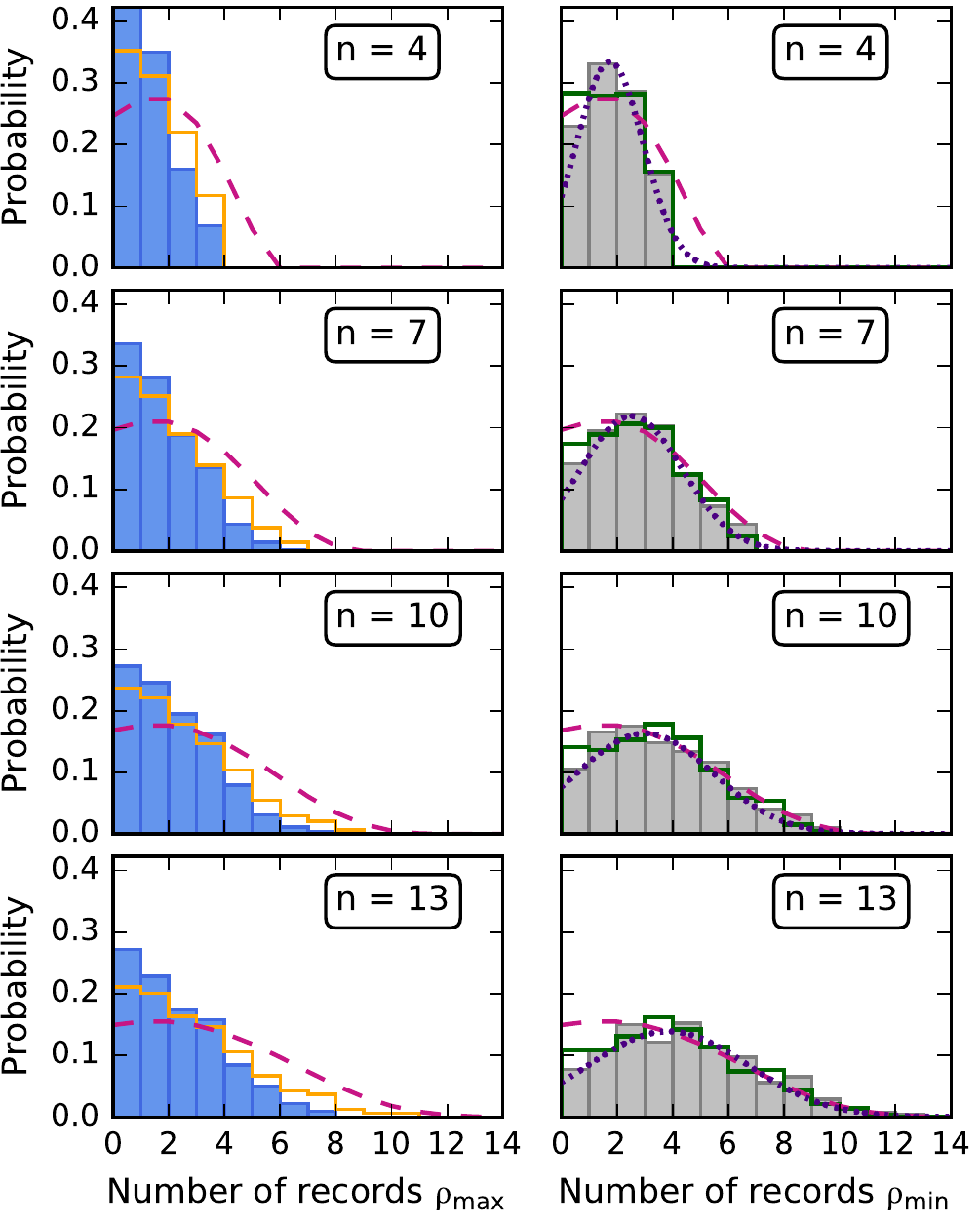}
\caption{Record number distributions shown for various trace length $n$ stated left. Experimental data for upper (blue, left) and lower (light gray, right) records is depicted. The solid lines are numerical simulations and the dashed line represents the analytic result for a system without drift Eq.~(\ref{eq:RecNumDist}). The dotted lines in the left panel indicate a Gaussian fit to the data according to Eq.~\eqref{eq:Gaussian} of which the constants $a(\mu)$ and $b(\mu)$ are extracted, for details see text.}
\label{fig:recNumDist}
\end{figure}
Beyond extreme values and record values, a more general quantity is the number of records occurring during a trace.
Based on the Sparre Andersen theorem one can calculate the probability distribution of having $\rho$ records in a trace of $n$ steps to be \cite{Majumdar2008,Majumdar2010b}
\begin{align}
\label{eq:RecNumDist}
P(\rho|n) = \binom{2n-\rho+1}{n}2^{-2n+\rho-1}.
\end{align}
with its first moment given by
\begin{align}
\label{eq:expectRecNum}
\langle \rho \rangle = (2n+1)\binom{2n}{n}2^{-2n}.
\end{align}
Both results are universal, therefore independent of the initial distance distribution, but only valid for systems without drift.
In Ref.~\cite{Majumdar2012} the authors have analytically derived the distribution of record numbers for various jump distance distributions for large step lengths $n$ in the case of a drift present. 
\begin{figure}
\includegraphics[width=0.46\textwidth]{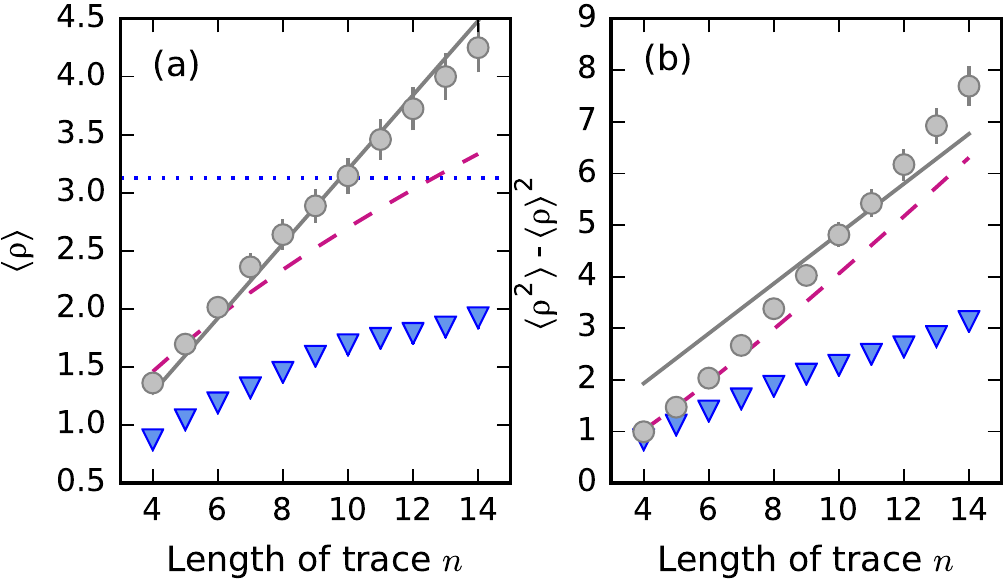}
\caption{Measured mean record number (a) and variance of the record number (b) for lower (light gray $\circ$) and upper (blue $\triangle$) records depicted for increasing trace length $n$. The dashed line indicates the theoretical expectation for systems without drift (cf. Eq.~\eqref{eq:expectRecNum}). The solid line depicts the expected values for $\langle \rho_\mathrm{min} \rangle$ and $\langle \rho_\mathrm{min}^2 \rangle-\langle \rho_\mathrm{min} \rangle^2$, respectively and the dotted line the theoretical saturation value for $\langle \rho_\mathrm{max} \rangle$. All three quantities are described by the constants $a(\mu)$ and $b(\mu)$, which are extracted from the data of Fig.~\ref{fig:recNumDist}, for details see text.
}
\label{fig:RecMoments}
\end{figure}
When $\phi(\xi)$ being exponential or Gaussian and $\mu \ge 0$ a Gaussian probability distribution for the upper record numbers has been found and is described by
\begin{equation}
\label{eq:Gaussian}
P(\rho_\mathrm{max}, n) = \frac{1}{\sqrt{2\pi b(\mu) n}}\exp\left(-\frac{(\rho_\mathrm{max}-a(\mu)n)^2}{2b(\mu)n}\right).
\end{equation}
Here $a(\mu)$ and $b(\mu)$ are constants depending only on the drift value $\mu$.
The resulting probability distributions for upper record numbers $\rho_\mathrm{max}$ and lower record numbers $\rho_\mathrm{min}$ as well as the data measured for $t_\mathrm{flight} = \SI{10}{\milli \s}$ is illustrated in Fig. \ref{fig:recNumDist}.
A clear difference between the measured probability distributions for $P(\rho_\mathrm{max})$ and $P(\rho_\mathrm{min})$ is visible: The upper records mainly preserve their shape, which is nearly exponential and exhibits a peak at $\rho = 0$ records. The distribution of lower record numbers has a more Gaussian shape and the most probable value is larger than one, in contrast to the case without drift, where zero stays the most probable value for all~$n$~\cite{Majumdar2010b}.

For systems with drift a linear increase of the mean lower record number $\langle \rho_\mathrm{min} \rangle \approx a(\mu)n$ and of the variance $\langle \rho^2_\mathrm{min} \rangle-\langle \rho_\mathrm{min} \rangle^2 \approx b(\mu) n$, has been calculated in \cite{Majumdar2012}, with $a(\mu)$ and $b(\mu)$ being the constants derived from Eq.~\eqref{eq:Gaussian}. Moreover, the first moment of the upper record number $\langle \rho_\mathrm{max} \rangle$ was predicted to show a saturation effect at the value of $1/a(\mu)$. 
We plot the first two moments in Fig.~\ref{fig:RecMoments} together with $a(\mu)$ and $b(\mu)$ describing the data very well. While the onset of saturation of $\langle \rho_\mathrm{max} \rangle$ can be seen in the data, the predicted value cannot be verified, due to the relatively small trace length.  

Thus, the recently predicted behaviour Eq.~\ref{eq:Gaussian} \cite{Majumdar2012} of record occurrence in drifting systems well describes our experimental data, particularly also the difference between upper and lower records. However, some information on the underlying distributions is necessary for the analysis.

\section{Conclusion}
We have demonstrated that implementation and analysis of individual atomic trajectories are an eminent and highly controllable testbed to study extreme value and record statistics in Markovian systems with drift in a broad range of parameters. In particular, employing single atoms diffusing on a periodic potential we have explored not only different widths of the distance distributions $\phi(\xi)$ or different strengths of drifts. We have rather also experimentally studied the case where single trajectories are not representative for the ensemble, and where the shape of the underlying Markov chain distributions change.
For such systems, the Sparre Andersen theorem holds, however, a first hint to the drift is already given by a reduced probability to cross the origin right in the beginning. 
We find that, the shape of the exponential distance distributions seems to dominate the distributions of extreme events. 
Moreover, we have studied the influence of a drift to four rare event distributions. We demonstrated that the drift can be extracted from extreme value statistics with the same precision as obtained from the position distribution, even if no knowledge about the underlying distributions is available. This may help, for example, in the case of large data sets where not the whole trace but only the extreme values need to be stored. 
In the distribution of occurrences of maxima and minima, which is a universal quantity for systems without drift, a clear sign of the drift is detected. The maxima are more likely to appear in the beginning, while the minima prevail at the end of the trace for the case of a negative drift. 
Finally, the record number distribution studied is again independent of the initial distance distribution $\phi(\xi)$. We confirmed that the lower record number distribution obeys a Gaussian distribution and furthermore we were able to show that both the mean value and the variance of the lower record number increases linearly with trace length as predicted in Ref.~\cite{Majumdar2012}. 
As the extreme value and record statistics are versatile tools to investigate Markov processes, we hope to stimulate further theoretical and experimental studies of this topic to analytically extend the experimental findings.  

\section*{Acknowledgements}
We thank E. Lutz for helpful discussions. The project was financially supported partially by the DFG via SFB/TR49 and partially by the European Union via the ERC Starting Grant 278208. T.L. acknowledges funding from Carl-Zeiss Stiftung, D.M. is a recipient of a DFG-fellowship through the Excellence Initiative by the Graduate School Materials Science in Mainz (GSC 266), and F.S. acknowledges funding by Studienstiftung des deutschen Volkes.

\bibliographystyle{apsrev4-1}
\bibliography{manuscript}

\begin{thebibliography}{35}%
\makeatletter
\providecommand \@ifxundefined [1]{%
 \@ifx{#1\undefined}
}%
\providecommand \@ifnum [1]{%
 \ifnum #1\expandafter \@firstoftwo
 \else \expandafter \@secondoftwo
 \fi
}%
\providecommand \@ifx [1]{%
 \ifx #1\expandafter \@firstoftwo
 \else \expandafter \@secondoftwo
 \fi
}%
\providecommand \natexlab [1]{#1}%
\providecommand \enquote  [1]{``#1''}%
\providecommand \bibnamefont  [1]{#1}%
\providecommand \bibfnamefont [1]{#1}%
\providecommand \citenamefont [1]{#1}%
\providecommand \href@noop [0]{\@secondoftwo}%
\providecommand \href [0]{\begingroup \@sanitize@url \@href}%
\providecommand \@href[1]{\@@startlink{#1}\@@href}%
\providecommand \@@href[1]{\endgroup#1\@@endlink}%
\providecommand \@sanitize@url [0]{\catcode `\\12\catcode `\$12\catcode
  `\&12\catcode `\#12\catcode `\^12\catcode `\_12\catcode `\%12\relax}%
\providecommand \@@startlink[1]{}%
\providecommand \@@endlink[0]{}%
\providecommand \url  [0]{\begingroup\@sanitize@url \@url }%
\providecommand \@url [1]{\endgroup\@href {#1}{\urlprefix }}%
\providecommand \urlprefix  [0]{URL }%
\providecommand \Eprint [0]{\href }%
\providecommand \doibase [0]{http://dx.doi.org/}%
\providecommand \selectlanguage [0]{\@gobble}%
\providecommand \bibinfo  [0]{\@secondoftwo}%
\providecommand \bibfield  [0]{\@secondoftwo}%
\providecommand \translation [1]{[#1]}%
\providecommand \BibitemOpen [0]{}%
\providecommand \bibitemStop [0]{}%
\providecommand \bibitemNoStop [0]{.\EOS\space}%
\providecommand \EOS [0]{\spacefactor3000\relax}%
\providecommand \BibitemShut  [1]{\csname bibitem#1\endcsname}%
\let\auto@bib@innerbib\@empty
\bibitem [{\citenamefont {Sabir}\ and\ \citenamefont
  {Santhanam}(2014)}]{Sabir2014}%
  \BibitemOpen
  \bibfield  {author} {\bibinfo {author} {\bibfnamefont {B.}~\bibnamefont
  {Sabir}}\ and\ \bibinfo {author} {\bibfnamefont {M.~S.}\ \bibnamefont
  {Santhanam}},\ }\href {\doibase 10.1103/PhysRevE.90.032126} {\bibfield
  {journal} {\bibinfo  {journal} {Phys. Rev. E - Stat. Nonlinear, Soft Matter
  Phys.}\ }\textbf {\bibinfo {volume} {90}},\ \bibinfo {pages} {032126}
  (\bibinfo {year} {2014})}\BibitemShut {NoStop}%
\bibitem [{\citenamefont {Redner}\ and\ \citenamefont
  {Petersen}(2006)}]{Redner2006}%
  \BibitemOpen
  \bibfield  {author} {\bibinfo {author} {\bibfnamefont {S.}~\bibnamefont
  {Redner}}\ and\ \bibinfo {author} {\bibfnamefont {M.~R.}\ \bibnamefont
  {Petersen}},\ }\href {\doibase 10.1103/PhysRevE.74.061114} {\bibfield
  {journal} {\bibinfo  {journal} {Phys. Rev. E - Stat. Nonlinear, Soft Matter
  Phys.}\ }\textbf {\bibinfo {volume} {74}},\ \bibinfo {pages} {061114}
  (\bibinfo {year} {2006})}\BibitemShut {NoStop}%
\bibitem [{\citenamefont {Redner}(2001)}]{Redner2001}%
  \BibitemOpen
  \bibfield  {author} {\bibinfo {author} {\bibfnamefont {S.}~\bibnamefont
  {Redner}},\ }\href@noop {} {\emph {\bibinfo {title} {A guide to first-passage
  processes}}}\ (\bibinfo  {publisher} {Cambridge University Press},\ \bibinfo
  {year} {2001})\BibitemShut {NoStop}%
\bibitem [{\citenamefont {Barry}\ \emph {et~al.}(1998)\citenamefont {Barry},
  \citenamefont {Balakrishnan},\ and\ \citenamefont {Nagaraja}}]{Barry1998}%
  \BibitemOpen
  \bibfield  {author} {\bibinfo {author} {\bibfnamefont {C.~A.}\ \bibnamefont
  {Barry}}, \bibinfo {author} {\bibfnamefont {N.}~\bibnamefont {Balakrishnan}},
  \ and\ \bibinfo {author} {\bibfnamefont {H.}~\bibnamefont {Nagaraja}},\
  }\href@noop {} {\emph {\bibinfo {title} {{Records}}}}\ (\bibinfo  {publisher}
  {John Wiley and Sons},\ \bibinfo {year} {1998})\BibitemShut {NoStop}%
\bibitem [{\citenamefont {Foster}\ and\ \citenamefont
  {Stuart}(1954)}]{Foster1954}%
  \BibitemOpen
  \bibfield  {author} {\bibinfo {author} {\bibfnamefont {F.~G.}\ \bibnamefont
  {Foster}}\ and\ \bibinfo {author} {\bibfnamefont {A.}~\bibnamefont
  {Stuart}},\ }\href@noop {} {\bibfield  {journal} {\bibinfo  {journal}
  {Journal of the Royal Statistical Society, Series B}\ }\textbf {\bibinfo
  {volume} {16}},\ \bibinfo {pages} {1} (\bibinfo {year} {1954})}\BibitemShut
  {NoStop}%
\bibitem [{\citenamefont {Glick}(1978)}]{Glick1978}%
  \BibitemOpen
  \bibfield  {author} {\bibinfo {author} {\bibfnamefont {N.}~\bibnamefont
  {Glick}},\ }\href@noop {} {\bibfield  {journal} {\bibinfo  {journal} {The
  American Mathematical Monthly}\ }\textbf {\bibinfo {volume} {85}},\ \bibinfo
  {pages} {2} (\bibinfo {year} {1978})}\BibitemShut {NoStop}%
\bibitem [{\citenamefont {Ben-Naim}\ \emph {et~al.}(2007)\citenamefont
  {Ben-Naim}, \citenamefont {Redner},\ and\ \citenamefont
  {Vazquez}}]{Ben-Naim2007}%
  \BibitemOpen
  \bibfield  {author} {\bibinfo {author} {\bibfnamefont {E.}~\bibnamefont
  {Ben-Naim}}, \bibinfo {author} {\bibfnamefont {S.}~\bibnamefont {Redner}}, \
  and\ \bibinfo {author} {\bibfnamefont {F.}~\bibnamefont {Vazquez}},\ }\href
  {\doibase 10.1209/0295-5075/77/30005} {\bibfield  {journal} {\bibinfo
  {journal} {Europhysics Letters (EPL)}\ }\textbf {\bibinfo {volume} {77}},\
  \bibinfo {pages} {30005} (\bibinfo {year} {2007})}\BibitemShut {NoStop}%
\bibitem [{\citenamefont {Gembris}\ \emph {et~al.}(2002)\citenamefont
  {Gembris}, \citenamefont {Taylor},\ and\ \citenamefont
  {Suter}}]{Gembris2002}%
  \BibitemOpen
  \bibfield  {author} {\bibinfo {author} {\bibfnamefont {D.}~\bibnamefont
  {Gembris}}, \bibinfo {author} {\bibfnamefont {J.~G.}\ \bibnamefont {Taylor}},
  \ and\ \bibinfo {author} {\bibfnamefont {D.}~\bibnamefont {Suter}},\ }\href
  {\doibase 10.1038/417506a} {\bibfield  {journal} {\bibinfo  {journal}
  {Nature}\ }\textbf {\bibinfo {volume} {417}},\ \bibinfo {pages} {506}
  (\bibinfo {year} {2002})}\BibitemShut {NoStop}%
\bibitem [{\citenamefont {Krug}(2007)}]{Krug2007}%
  \BibitemOpen
  \bibfield  {author} {\bibinfo {author} {\bibfnamefont {J.}~\bibnamefont
  {Krug}},\ }\href {http://stacks.iop.org/1742-5468/2007/i=07/a=P07001}
  {\bibfield  {journal} {\bibinfo  {journal} {Journal of Statistical Mechanics:
  Theory and Experiment}\ }\textbf {\bibinfo {volume} {2007}},\ \bibinfo
  {pages} {P07001} (\bibinfo {year} {2007})}\BibitemShut {NoStop}%
\bibitem [{\citenamefont {Sibani}\ \emph {et~al.}(2006)\citenamefont {Sibani},
  \citenamefont {Rodriguez},\ and\ \citenamefont {Kenning}}]{Sibani2006}%
  \BibitemOpen
  \bibfield  {author} {\bibinfo {author} {\bibfnamefont {P.}~\bibnamefont
  {Sibani}}, \bibinfo {author} {\bibfnamefont {G.~F.}\ \bibnamefont
  {Rodriguez}}, \ and\ \bibinfo {author} {\bibfnamefont {G.~G.}\ \bibnamefont
  {Kenning}},\ }\href {\doibase 10.1103/PhysRevB.74.224407} {\bibfield
  {journal} {\bibinfo  {journal} {Physical Review B}\ }\textbf {\bibinfo
  {volume} {74}},\ \bibinfo {pages} {224407} (\bibinfo {year}
  {2006})}\BibitemShut {NoStop}%
\bibitem [{\citenamefont {Bhosale}\ \emph {et~al.}(2012)\citenamefont
  {Bhosale}, \citenamefont {Tomsovic},\ and\ \citenamefont
  {Lakshminarayan}}]{Bhosale2012}%
  \BibitemOpen
  \bibfield  {author} {\bibinfo {author} {\bibfnamefont {U.~T.}\ \bibnamefont
  {Bhosale}}, \bibinfo {author} {\bibfnamefont {S.}~\bibnamefont {Tomsovic}}, \
  and\ \bibinfo {author} {\bibfnamefont {A.}~\bibnamefont {Lakshminarayan}},\
  }\href {\doibase 10.1103/PhysRevA.85.062331} {\bibfield  {journal} {\bibinfo
  {journal} {Phys. Rev. A}\ }\textbf {\bibinfo {volume} {85}},\ \bibinfo
  {pages} {062331} (\bibinfo {year} {2012})}\BibitemShut {NoStop}%
\bibitem [{\citenamefont {{Srivastava, Shashi C. L.}}\ \emph
  {et~al.}(2013)\citenamefont {{Srivastava, Shashi C. L.}}, \citenamefont
  {{Lakshminarayan, Arul}},\ and\ \citenamefont {{Jain, Sudhir
  R.}}}]{Shashi2013}%
  \BibitemOpen
  \bibfield  {author} {\bibinfo {author} {\bibnamefont {{Srivastava, Shashi C.
  L.}}}, \bibinfo {author} {\bibnamefont {{Lakshminarayan, Arul}}}, \ and\
  \bibinfo {author} {\bibnamefont {{Jain, Sudhir R.}}},\ }\href {\doibase
  10.1209/0295-5075/101/10003} {\bibfield  {journal} {\bibinfo  {journal}
  {EPL}\ }\textbf {\bibinfo {volume} {101}},\ \bibinfo {pages} {10003}
  (\bibinfo {year} {2013})}\BibitemShut {NoStop}%
\bibitem [{\citenamefont {Denk}\ and\ \citenamefont {Webb}(1989)}]{Denk1989}%
  \BibitemOpen
  \bibfield  {author} {\bibinfo {author} {\bibfnamefont {W.}~\bibnamefont
  {Denk}}\ and\ \bibinfo {author} {\bibfnamefont {W.~W.}\ \bibnamefont
  {Webb}},\ }\href {\doibase 10.1103/PhysRevLett.63.207} {\bibfield  {journal}
  {\bibinfo  {journal} {Physical Review Letters}\ }\textbf {\bibinfo {volume}
  {63}},\ \bibinfo {pages} {207} (\bibinfo {year} {1989})}\BibitemShut
  {NoStop}%
\bibitem [{\citenamefont {Schweitzer}\ and\ \citenamefont {{Doyne
  Farmer}}(2007)}]{SchweitzerFrank2007}%
  \BibitemOpen
  \bibfield  {author} {\bibinfo {author} {\bibfnamefont {F.}~\bibnamefont
  {Schweitzer}}\ and\ \bibinfo {author} {\bibfnamefont {J.}~\bibnamefont
  {{Doyne Farmer}}},\ }\href@noop {} {\emph {\bibinfo {title} {{Brownian agents
  and active particles: collective dynamics in the natural and social
  sciences}}}}\ (\bibinfo  {publisher} {Springer Science {\&} Business Media},\
  \bibinfo {year} {2007})\BibitemShut {NoStop}%
\bibitem [{\citenamefont {van~den Engh}\ \emph {et~al.}(1992)\citenamefont
  {van~den Engh}, \citenamefont {Sachs},\ and\ \citenamefont
  {Trask}}]{vandenEngh1992}%
  \BibitemOpen
  \bibfield  {author} {\bibinfo {author} {\bibfnamefont {G.}~\bibnamefont
  {van~den Engh}}, \bibinfo {author} {\bibfnamefont {R.}~\bibnamefont {Sachs}},
  \ and\ \bibinfo {author} {\bibfnamefont {B.~J.}\ \bibnamefont {Trask}},\
  }\href {\doibase 10.1126/science.1388286} {\bibfield  {journal} {\bibinfo
  {journal} {Science (New York, N.Y.)}\ }\textbf {\bibinfo {volume} {257}},\
  \bibinfo {pages} {1410} (\bibinfo {year} {1992})}\BibitemShut {NoStop}%
\bibitem [{\citenamefont {Bouchaud}\ and\ \citenamefont
  {Potters}(2003)}]{Bouchaud2003}%
  \BibitemOpen
  \bibfield  {author} {\bibinfo {author} {\bibfnamefont {J.-P.}\ \bibnamefont
  {Bouchaud}}\ and\ \bibinfo {author} {\bibfnamefont {M.}~\bibnamefont
  {Potters}},\ }\href@noop {} {\emph {\bibinfo {title} {{Theory of financial
  risk and derivative pricing: from statistical physics to risk management}}}}\
  (\bibinfo  {publisher} {Cambridge University Press},\ \bibinfo {year}
  {2003})\BibitemShut {NoStop}%
\bibitem [{\citenamefont {Sabhapandit}(2011)}]{Sabhapandit2011}%
  \BibitemOpen
  \bibfield  {author} {\bibinfo {author} {\bibfnamefont {S.}~\bibnamefont
  {Sabhapandit}},\ }\href {\doibase 10.1209/0295-5075/94/20003} {\bibfield
  {journal} {\bibinfo  {journal} {EPL (Europhysics Letters)}\ }\textbf
  {\bibinfo {volume} {94}},\ \bibinfo {pages} {20003} (\bibinfo {year}
  {2011})}\BibitemShut {NoStop}%
\bibitem [{\citenamefont {Majumdar}(2010)}]{Majumdar2010b}%
  \BibitemOpen
  \bibfield  {author} {\bibinfo {author} {\bibfnamefont {S.~N.}\ \bibnamefont
  {Majumdar}},\ }\href {\doibase 10.1016/j.physa.2010.01.021} {\bibfield
  {journal} {\bibinfo  {journal} {Physica A: Statistical Mechanics and its
  Applications}\ }\textbf {\bibinfo {volume} {389}},\ \bibinfo {pages} {4299}
  (\bibinfo {year} {2010})}\BibitemShut {NoStop}%
\bibitem [{\citenamefont {{Le Doussal}}\ and\ \citenamefont
  {Wiese}(2009)}]{LeDoussal2009}%
  \BibitemOpen
  \bibfield  {author} {\bibinfo {author} {\bibfnamefont {P.}~\bibnamefont {{Le
  Doussal}}}\ and\ \bibinfo {author} {\bibfnamefont {K.~J.}\ \bibnamefont
  {Wiese}},\ }\href {\doibase 10.1103/PhysRevE.79.051105} {\bibfield  {journal}
  {\bibinfo  {journal} {Physical Review E - Statistical, Nonlinear, and Soft
  Matter Physics}\ }\textbf {\bibinfo {volume} {79}},\ \bibinfo {pages}
  {051105} (\bibinfo {year} {2009})}\BibitemShut {NoStop}%
\bibitem [{\citenamefont {Majumdar}\ \emph {et~al.}(2012)\citenamefont
  {Majumdar}, \citenamefont {Schehr},\ and\ \citenamefont
  {Wergen}}]{Majumdar2012}%
  \BibitemOpen
  \bibfield  {author} {\bibinfo {author} {\bibfnamefont {S.~N.}\ \bibnamefont
  {Majumdar}}, \bibinfo {author} {\bibfnamefont {G.}~\bibnamefont {Schehr}}, \
  and\ \bibinfo {author} {\bibfnamefont {G.}~\bibnamefont {Wergen}},\ }\href
  {\doibase 10.1088/1751-8113/45/35/355002} {\bibfield  {journal} {\bibinfo
  {journal} {Journal of Physics A: Mathematical and Theoretical}\ }\textbf
  {\bibinfo {volume} {45}},\ \bibinfo {pages} {1} (\bibinfo {year}
  {2012})}\BibitemShut {NoStop}%
\bibitem [{\citenamefont {Gerber}\ and\ \citenamefont
  {Shiu}(1996)}]{Gerber1996}%
  \BibitemOpen
  \bibfield  {author} {\bibinfo {author} {\bibfnamefont {H.~U.}\ \bibnamefont
  {Gerber}}\ and\ \bibinfo {author} {\bibfnamefont {E.~S.~W.}\ \bibnamefont
  {Shiu}},\ }\href {\doibase 10.1016/0167-6687(96)85007-4} {\bibfield
  {journal} {\bibinfo  {journal} {Insurance: Mathematics and Economics}\
  }\textbf {\bibinfo {volume} {18}},\ \bibinfo {pages} {183} (\bibinfo {year}
  {1996})}\BibitemShut {NoStop}%
\bibitem [{\citenamefont {Dickson}(2001)}]{Dickson2001}%
  \BibitemOpen
  \bibfield  {author} {\bibinfo {author} {\bibfnamefont {D.}~\bibnamefont
  {Dickson}},\ }\href
  {http://www.sciencedirect.com/science/article/pii/S0167668701000919}
  {\bibfield  {journal} {\bibinfo  {journal} {Insurance: Mathematics and
  Economics}\ }\textbf {\bibinfo {volume} {29}},\ \bibinfo {pages} {333}
  (\bibinfo {year} {2001})}\BibitemShut {NoStop}%
\bibitem [{\citenamefont {Villarroel}\ and\ \citenamefont
  {Montero}(2010)}]{Villarroel2010}%
  \BibitemOpen
  \bibfield  {author} {\bibinfo {author} {\bibfnamefont {J.}~\bibnamefont
  {Villarroel}}\ and\ \bibinfo {author} {\bibfnamefont {M.}~\bibnamefont
  {Montero}},\ }\href {\doibase 10.1088/0953-4075/43/13/135404} {\bibfield
  {journal} {\bibinfo  {journal} {Journal of Physics B: Atomic, Molecular and
  Optical Physics}\ }\textbf {\bibinfo {volume} {43}},\ \bibinfo {pages}
  {135404} (\bibinfo {year} {2010})}\BibitemShut {NoStop}%
\bibitem [{\citenamefont {Edery}\ \emph {et~al.}(2011)\citenamefont {Edery},
  \citenamefont {Kostinski},\ and\ \citenamefont {Berkowitz}}]{Edery2011}%
  \BibitemOpen
  \bibfield  {author} {\bibinfo {author} {\bibfnamefont {Y.}~\bibnamefont
  {Edery}}, \bibinfo {author} {\bibfnamefont {A.}~\bibnamefont {Kostinski}}, \
  and\ \bibinfo {author} {\bibfnamefont {B.}~\bibnamefont {Berkowitz}},\ }\href
  {\doibase 10.1029/2011GL048558} {\bibfield  {journal} {\bibinfo  {journal}
  {Geophysical Research Letters}\ }\textbf {\bibinfo {volume} {38}},\ \bibinfo
  {pages} {L16403} (\bibinfo {year} {2011})}\BibitemShut {NoStop}%
\bibitem [{\citenamefont {Kindermann}\ \emph {et~al.}(2017)\citenamefont
  {Kindermann}, \citenamefont {Dechant}, \citenamefont {Hohmann}, \citenamefont
  {Lausch}, \citenamefont {Mayer}, \citenamefont {Schmidt}, \citenamefont
  {Lutz},\ and\ \citenamefont {Widera}}]{Kindermann2016}%
  \BibitemOpen
  \bibfield  {author} {\bibinfo {author} {\bibfnamefont {F.}~\bibnamefont
  {Kindermann}}, \bibinfo {author} {\bibfnamefont {A.}~\bibnamefont {Dechant}},
  \bibinfo {author} {\bibfnamefont {M.}~\bibnamefont {Hohmann}}, \bibinfo
  {author} {\bibfnamefont {T.}~\bibnamefont {Lausch}}, \bibinfo {author}
  {\bibfnamefont {D.}~\bibnamefont {Mayer}}, \bibinfo {author} {\bibfnamefont
  {F.}~\bibnamefont {Schmidt}}, \bibinfo {author} {\bibfnamefont
  {E.}~\bibnamefont {Lutz}}, \ and\ \bibinfo {author} {\bibfnamefont
  {A.}~\bibnamefont {Widera}},\ }\href@noop {} {\bibfield  {journal} {\bibinfo
  {journal} {Nature Physics}\ }\textbf {\bibinfo {volume} {13}},\ \bibinfo
  {pages} {137} (\bibinfo {year} {2017})}\BibitemShut {NoStop}%
\bibitem [{\citenamefont {Edery}\ \emph {et~al.}(2013)\citenamefont {Edery},
  \citenamefont {Kostinski}, \citenamefont {Majumdar},\ and\ \citenamefont
  {Berkowitz}}]{Edery2013}%
  \BibitemOpen
  \bibfield  {author} {\bibinfo {author} {\bibfnamefont {Y.}~\bibnamefont
  {Edery}}, \bibinfo {author} {\bibfnamefont {A.~B.}\ \bibnamefont
  {Kostinski}}, \bibinfo {author} {\bibfnamefont {S.~N.}\ \bibnamefont
  {Majumdar}}, \ and\ \bibinfo {author} {\bibfnamefont {B.}~\bibnamefont
  {Berkowitz}},\ }\href {\doibase 10.1103/PhysRevLett.110.180602} {\bibfield
  {journal} {\bibinfo  {journal} {Physical Review Letters}\ }\textbf {\bibinfo
  {volume} {110}},\ \bibinfo {pages} {180602} (\bibinfo {year}
  {2013})}\BibitemShut {NoStop}%
\bibitem [{\citenamefont {Steck}(2008)}]{Steck2008}%
  \BibitemOpen
  \bibfield  {author} {\bibinfo {author} {\bibfnamefont {D.}~\bibnamefont
  {Steck}},\ }\href {http://steck.us/alkalidata/sodiumnumbers.pdf} {\enquote
  {\bibinfo {title} {{Cesium D Line Data}},}\ } (\bibinfo {year}
  {2008})\BibitemShut {NoStop}%
\bibitem [{\citenamefont {H{\"{a}}nggi}\ \emph {et~al.}(1990)\citenamefont
  {H{\"{a}}nggi}, \citenamefont {Talkner},\ and\ \citenamefont
  {Borkovec}}]{Hanggi1990}%
  \BibitemOpen
  \bibfield  {author} {\bibinfo {author} {\bibfnamefont {P.}~\bibnamefont
  {H{\"{a}}nggi}}, \bibinfo {author} {\bibfnamefont {P.}~\bibnamefont
  {Talkner}}, \ and\ \bibinfo {author} {\bibfnamefont {M.}~\bibnamefont
  {Borkovec}},\ }\href {\doibase 10.1103/RevModPhys.62.251} {\bibfield
  {journal} {\bibinfo  {journal} {Rev. Mod. Phys.}\ }\textbf {\bibinfo {volume}
  {62}},\ \bibinfo {pages} {251} (\bibinfo {year} {1990})}\BibitemShut
  {NoStop}%
\bibitem [{Note1()}]{Note1}%
  \BibitemOpen
  \bibinfo {note} {The limited width of the detection region also contributes
  to the complex drift observed.}\BibitemShut {Stop}%
\bibitem [{\citenamefont {{Sparre Andersen}}(1954)}]{SparreAndersen1954}%
  \BibitemOpen
  \bibfield  {author} {\bibinfo {author} {\bibfnamefont {E.}~\bibnamefont
  {{Sparre Andersen}}},\ }\href@noop {} {\bibfield  {journal} {\bibinfo
  {journal} {Math. Scand.}\ }\textbf {\bibinfo {volume} {2}},\ \bibinfo {pages}
  {195} (\bibinfo {year} {1954})}\BibitemShut {NoStop}%
\bibitem [{\citenamefont {Pollaczek}(1952)}]{Pollaczek1952}%
  \BibitemOpen
  \bibfield  {author} {\bibinfo {author} {\bibfnamefont {F.}~\bibnamefont
  {Pollaczek}},\ }\href@noop {} {\bibfield  {journal} {\bibinfo  {journal}
  {Comptes Rendus Hebdomadaires des Seances de L’Academie Des Sciences}\
  }\textbf {\bibinfo {volume} {234}},\ \bibinfo {pages} {2334} (\bibinfo {year}
  {1952})}\BibitemShut {NoStop}%
\bibitem [{\citenamefont {Spitzer}(1956)}]{Spitzer1956}%
  \BibitemOpen
  \bibfield  {author} {\bibinfo {author} {\bibfnamefont {F.}~\bibnamefont
  {Spitzer}},\ }\href {\doibase 10.1090/S0002-9947-1956-0079851-X} {\bibfield
  {journal} {\bibinfo  {journal} {Transactions of the American Mathematical
  Society}\ }\textbf {\bibinfo {volume} {82}},\ \bibinfo {pages} {323}
  (\bibinfo {year} {1956})}\BibitemShut {NoStop}%
\bibitem [{\citenamefont {Majumdar}(2003)}]{Majumdar2003}%
  \BibitemOpen
  \bibfield  {author} {\bibinfo {author} {\bibfnamefont {S.}~\bibnamefont
  {Majumdar}},\ }\href {\doibase 10.1016/S0378-4371(02)01422-X} {\bibfield
  {journal} {\bibinfo  {journal} {Physica A: Statistical Mechanics and its
  Applications}\ }\textbf {\bibinfo {volume} {318}},\ \bibinfo {pages} {161}
  (\bibinfo {year} {2003})}\BibitemShut {NoStop}%
\bibitem [{\citenamefont {Comtet}\ and\ \citenamefont
  {Majumdar}(2005)}]{Comtet2005}%
  \BibitemOpen
  \bibfield  {author} {\bibinfo {author} {\bibfnamefont {A.}~\bibnamefont
  {Comtet}}\ and\ \bibinfo {author} {\bibfnamefont {S.~N.}\ \bibnamefont
  {Majumdar}},\ }\href {\doibase 10.1088/1742-5468/2005/06/P06013} {\bibfield
  {journal} {\bibinfo  {journal} {J. Stat. Mech. Theory Exp.}\ }\textbf
  {\bibinfo {volume} {2005}},\ \bibinfo {pages} {P06013} (\bibinfo {year}
  {2005})}\BibitemShut {NoStop}%
\bibitem [{\citenamefont {Majumdar}\ and\ \citenamefont
  {Ziff}(2008)}]{Majumdar2008}%
  \BibitemOpen
  \bibfield  {author} {\bibinfo {author} {\bibfnamefont {S.~N.}\ \bibnamefont
  {Majumdar}}\ and\ \bibinfo {author} {\bibfnamefont {R.~M.}\ \bibnamefont
  {Ziff}},\ }\href {\doibase 10.1103/PhysRevLett.101.050601} {\bibfield
  {journal} {\bibinfo  {journal} {Physical Review Letters}\ }\textbf {\bibinfo
  {volume} {101}},\ \bibinfo {pages} {050601} (\bibinfo {year}
  {2008})}\BibitemShut {NoStop}%
\end{thebibliography}%

\end{document}